\documentclass[reprint,twocolumn,amsmath,amssymb,aps,prl,superscriptaddress]{revtex4-1}

\usepackage{graphicx}
\usepackage{subfigure}
\usepackage{dcolumn}
\usepackage{bm}
\begin{document}

\title{Search for invisible axion dark matter with a multiple-cell haloscope}

\author{Junu Jeong}
\affiliation{Center for Axion and Precision Physics Research, IBS, Daejeon 34051 Republic of Korea}
\affiliation{Department of Physics, KAIST, Daejeon 34141 Republic of Korea}
\author{SungWoo Youn}
\email{Corresponding author (swyoun@ibs.re.kr)}
\affiliation{Center for Axion and Precision Physics Research, IBS, Daejeon 34051 Republic of Korea}
\author{Sungjae Bae}
\affiliation{Center for Axion and Precision Physics Research, IBS, Daejeon 34051 Republic of Korea}
\affiliation{Department of Physics, KAIST, Daejeon 34141 Republic of Korea}
\author{Jihngeun Kim}
\email{Now at SuperGenics Co., Ltd., Changwon 51542 Republic of Korea}
\author{Taehyeon Seong}
\affiliation{Center for Axion and Precision Physics Research, IBS, Daejeon 34051 Republic of Korea}
\author{Jihn E Kim}
\affiliation{Department of Physics, Kyung Hee University, Seoul 02447 Republic of Korea}
\author{Yannis K. Semertzidis}
\affiliation{Center for Axion and Precision Physics Research, IBS, Daejeon 34051 Republic of Korea}
\affiliation{Department of Physics, KAIST, Daejeon 34141 Republic of Korea}

\date{\today}

\begin{abstract}
We present the first results of a search for invisible axion dark matter using a multiple-cell cavity haloscope.
This cavity concept was proposed to provide a highly efficient approach to high mass regions compared to the conventional multiple-cavity design, with larger detection volume, simpler detector setup, and unique phase-matching mechanism.
Searches with a double-cell cavity superseded previous reports for the axion-photon coupling over the mass range between 13.0 and 13.9\,$\mu$eV.
This result not only demonstrates the novelty of the cavity concept for high-mass axion searches, but also suggests it can make considerable contributions to the next-generation experiments.
\end{abstract}

%\keywords{Suggested keywords}%Use showkeys class option if keyword
                              %display desired
\maketitle

The Peccei-Quinn mechanism was proposed in the 1970s as a dynamic solution to the CP symmetry problem in quantum chromodynamics (QCD)~\cite{paper:PQ}. 
It involves a new global U(1) symmetry whose spontaneous breaking gives rise to a pseudo-Nambu-Goldstone boson - the axion~\cite{paper:axion}.
The null results from searches for standard axions have led to the attention of very light (``invisible") axions~\cite{paper:invisible}.
Axions could also account for the halos of dark matter, the mysterious substance that is believed to constitute 85\% of the matter in the universe~\cite{paper:CDM}.
The interesting mass range is constrained by cosmological constraints and astrophysical observations~\cite{paper:SN1987A}.
The axion has well-defined properties including 1) its mass, which is determined by the energy scale associated with global symmetry breaking; and 2) its coupling to standard model particles depending on theoretical models $-$ KSVZ~\cite{paper:invisible, paper:KSVZ} and DFSZ~\cite{paper:DFSZ}.

A promising detection scheme relies on the axion-to-photon conversion provoked by virtual photons generated by a magnetic field~\cite{paper:Sikivie}.
The cavity haloscope is one of the most sensitive approaches to investigating QCD axion physics in the $\mu$eV mass range. 
It employs a microwave resonator in a strong magnetic field to enhance the photon signal power~\cite{paper:detection_rate}
\begin{equation}
P_{a\rightarrow\gamma\gamma}=g_{a\gamma\gamma}^2\frac{\rho_a}{m_a}B_0^2VC\frac{Q_cQ_a}{Q_c+Q_a},
\label{eq:signal}
\end{equation}
where $g_{a\gamma\gamma}$ is the axion-to-photon coupling, $\rho_a$ is the local axion dark matter halo density, $m_a$ is the axion mass, $B_0$ is the external magnetic field, $V$ is the cavity volume, $C$ is the geometry factor of the resonant mode under consideration, and $Q_c$ and $Q_a$ are the cavity and axion quality factors, respectively.
Experimental sensitivity is determined by the signal-to-noise ratio (SNR), where the noise is described as fluctuations in system noise power obtained over a bandwidth $\Delta\nu$ within a time $\Delta t$, i.e.,
\begin{equation}
\delta P_{\rm sys}=k_BT_{\rm sys}\sqrt{\frac{\Delta\nu}{\Delta t}},
\label{eq:noise}
\end{equation}
using the Boltzmann constant $k_B$ and the system noise temperature $T_{\rm sys}$~\cite{paper:radiometer, paper:johnson-nyquist}.
$T_{\rm sys}$ is given by a linear combination of the cavity temperature and the equivalent noise temperature of the receiver chain.
Since the axion mass is {\it a priori} unknown, experiments concentrate on how fast possible mass ranges can be scanned.
The relevant quantity is called the scan rate, which is formulated as~\cite{paper:detection_rate}
\begin{equation}
\frac{d\nu}{dt} = g_{a\gamma\gamma}^4 \frac{\rho_a^2} {m_a^2} \frac{1}{\rm SNR^2} \frac{B_0^4V^2C^2 Q_c Q_a}{k_B^2T_{\rm sys}^2} \frac{\beta^2}{(1+\beta)^3},
\label{eq:scan_rate}
\end{equation}
where $\beta$ denotes the coupling strength of a receiver and a typical assumption of $Q_c \ll Q_a$ is adopted.

Many experimental efforts have been conducted to exploit this technique, relying on the fundamental TM$_{010}$ mode of a cylindrical resonator placed in a superconducting (SC) solenoid.
However, they are sensitive to relatively low mass regions $(<10\,\mu{\rm eV})$ since a single cavity is employed to make the most of the given volume, such as ADMX~\cite{paper:ADMX} and CAPP-8TB~\cite{paper:CAPP-8TB}.
Higher-mass axion searches have also been demonstrated, yet their sensitivities were limited mainly because of compromises between the resonant frequency and volume of the cavity -- HAYSTAC~\cite{paper:HAYSTAC}, ORGAN~\cite{paper:ORGAN}, ADMX~\cite{paper:sidecar}, and QUAX-a$\gamma$~\cite{paper:QUAX-ar}.
Meanwhile, several studies have been performed on cavity design to compensate for the volume loss, e.g., arrays of multiple cavities~\cite{thesis:ADMX_multicav, paper:multiple_cavity}, cavities with equidistant partitions~\cite{paper:multiple_cell, paper:RADES}, designs exploiting higher-order modes~\cite{paper:supermode, paper:wheel, paper:DBAS}, and high-$Q$ cavities~\cite{paper:SC_cavity, paper:QUAX}.
In this Letter, we apply a novel cavity design, known as multiple-cell cavity, to high-mass axion searches and report the first results using a double-cell cavity haloscope.

\begin{figure}
    \centering
    \includegraphics[width=\linewidth]{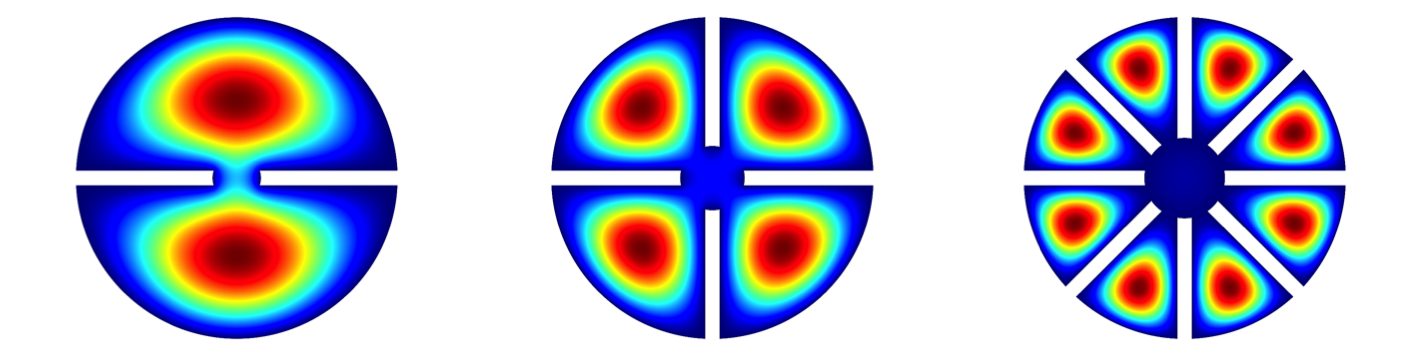}
    \caption{Various multiple-cell (double-, quadruple- and octuple-cell) cavities with the E-field distributions of the TM$_{010}$-like mode.}
    \label{fig:pizza_cavities}
\end{figure}

The multiple-cell cavity design, characterized by conducting partitions which vertically divide the cavity volume into identical cells, effectively increases the resonant frequencies with minimal volume loss, as illustrated in Fig.~\ref{fig:pizza_cavities}.
The gap between partitions in the middle of the cavity plays several critical roles:
1) it spatially connects all the cells, enabling a single antenna to pick up the signal from the entire volume and thus significantly simplifying the structure of the receiver chain; 
2) it breaks the frequency degeneracy with the lowest mode corresponding to the TM$_{010}$-like mode regardless of the cell multiplicity, making the mode identification straightforward; and
3) it vanishes the E-field of the degeneracy-broken higher modes at the center of the cavity under ideal circumstances, imposing a unique condition to ensure the field is evenly distributed, and thus maximize the effective detection volume~\cite{paper:multiple_cell}.

\begin{figure}[h]
    \centering
    \includegraphics[width=0.8\linewidth]{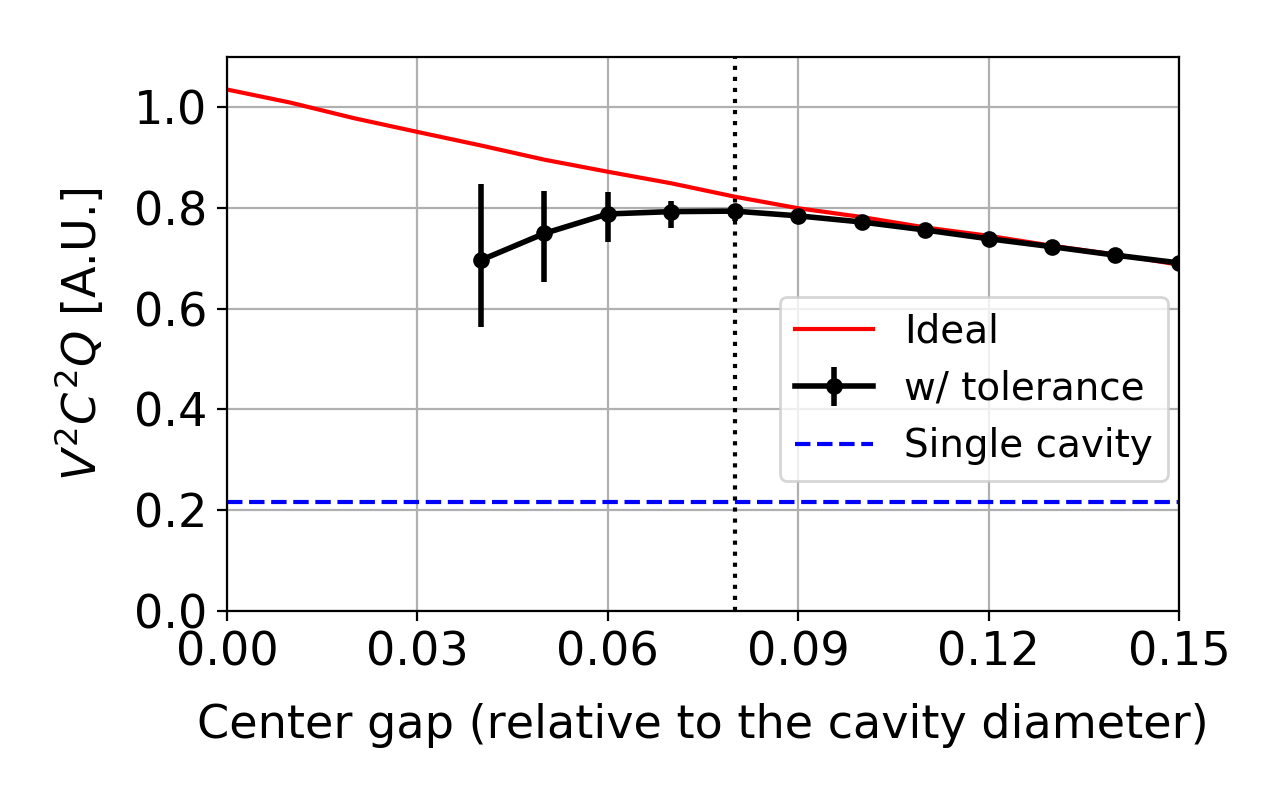}
    \caption{Effects of machining tolerance from a 2D double-cell cavity model, represented by the points with error bars, denoting the systematic uncertainties from the simulation, as compared with the ideal situation.
    The optimal gap is indicated by the vertical dotted line.
    The performance of a standard single cavity geometry yielding the same resonant frequency is contrasted.}
    \label{fig:gap_opt}
\end{figure}

In reality, however, dimensional tolerances in cavity construction can cause the EM field to be localized in particular cells, which eventually reduces the effective volume.
To estimate the effect, a 2-dimensional simulation study was conducted using the COMSOL Multiphysics software~\cite{tool:comsol}.
Starting with an ideal cavity model, we varied several dimensions -- the radius of each cell and the thickness and position of each partition -- independently by the amounts randomly selected from a normal distribution with $\sigma=100\,\mu$m, which corresponds to the typical machining tolerance~\cite{standard:IT_grade}.
For a fixed partition gap, such perturbations were modelled 100 times, and in each case a quantity $V^2C^2Q$, a product of cavity-associated parameters in Eq.~\ref{eq:scan_rate}, was calculated.
This procedure was repeated for differently sized center gap.
The simulation result for a double-cell cavity is shown in Fig.~\ref{fig:gap_opt}, where we notice that there is an optimal gap to minimize the effect.

To avoid complications in tuning system with increasing cell multiplicity, the tuning mechanism of Ref.~\cite{paper:multiple_cell} was revised.
To the top and bottom of each cell, we introduced an arch-shaped opening through which a tuning rod can be extended out of the cavity.
The opening stretches from a partition to the cell field center to achieve large frequency tunability.
The set of tuning rods are gripped at both ends by a pair of G-10 bars outside the cavity:
beneath the cavity, a 6mm-thick bar is mounted on a single rotational actuator, which eventually moves all the rods simultaneously, while on the cavity top, the other bar with a hole in the middle through which a pick-up antenna can be inserted into the cavity, guides them for symmetric motion with respect to the cavity center.

\begin{figure}
    \centering
    \includegraphics[width=0.8\linewidth]{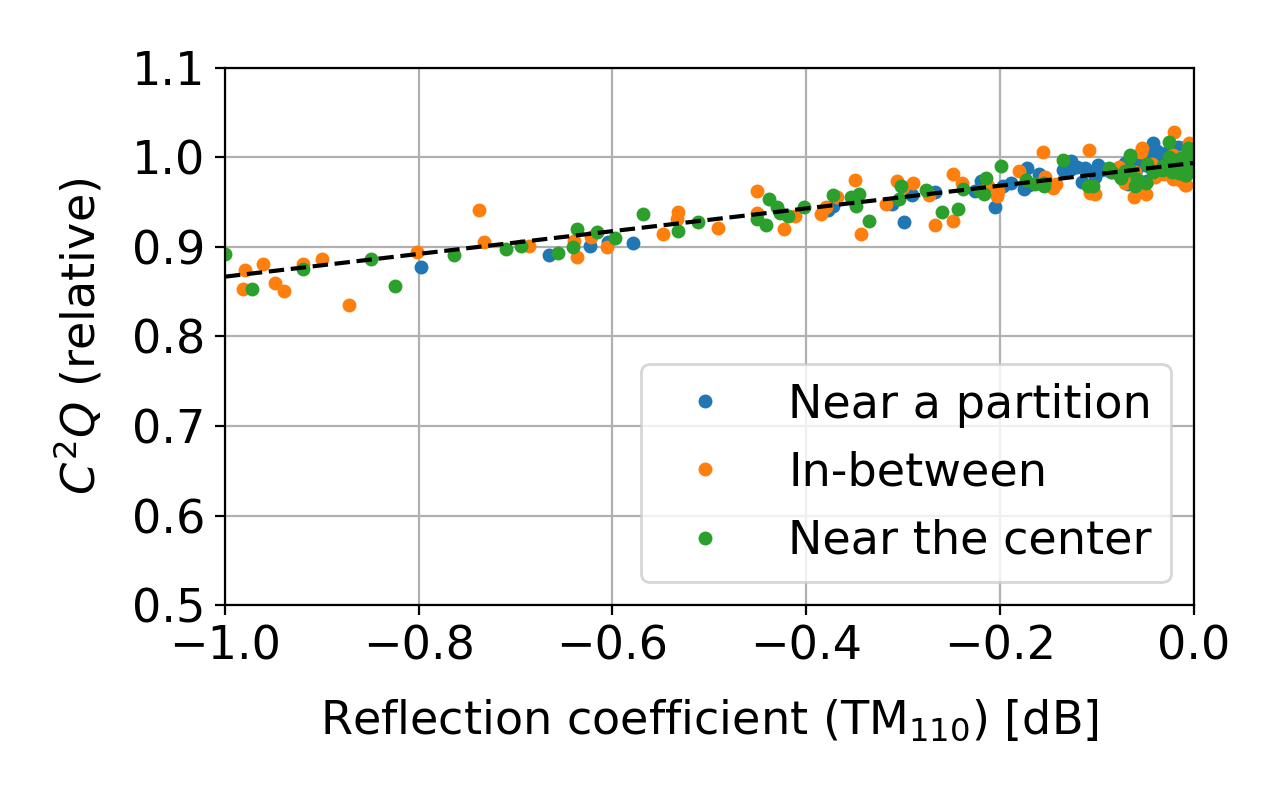}
    \caption{Relative scan rate vs. reflection coefficient for the TM$_{110}$-like mode probed at the cavity center at different rod positions.}
    \label{fig:misalignment}
\end{figure}

Concerning the potential field localization due to misalignment of the tuning system, the effect was also estimated based on a simulation study. 
For simplicity, we modelled a double-cell cavity system with the optimal partition gap and a pair of identical dielectric rods with $\epsilon_r=10$.
The pair was artificially misaligned so that the EM field symmetry was slightly broken, causing the field to populate one cell more than the other.
Such field localization is characterized by a non-vanishing E-field of the higher (TM$_{110}$-like) mode in the cavity center, which can be translated into the non-zero coupling strength, an experimentally measurable quantity.
For each misalignment, the reflection coefficient was computed and a cavity-associated quantity $C^2Q$ was calculated.
This procedure was repeated at several different positions along the arch-shaped holes and the result is shown in Fig.~\ref{fig:misalignment}.
Typically measured values $< 0.05$\,dB correspond to $< 1\%$ reductions in scan rate. 

An axion search experiment employing a double-cell cavity was carried out utilizing a He-3 cryogenic system and a 9T SC magnet, after which the experiment was named ``CAPP-9T".
The cryogenic system, containing 10 liters of He-3 gas, employs a set of pumps: 1) a 1K pot pump to condense the gas and collect the liquid in the He-3 pot, and (2) a charcoal sorption pump to adjust the vapor pressure and thus control the temperature of the He-3 plate.
System operation at the minimum vapor pressure maintained the base temperature of the He-3 plate at $\sim300$\,mK ($\sim700$\,mK) for about 200 (1) hours without a load (with a load of $\sim10$\,kg).
We chose an operating mode with a high vapor pressure by keeping the charcoal temperature at $\sim40$\,K, because it continuously maintained the He-3 plate (cavity) temperature at about 1.7\,K (2.1\,K).
The 9T SC magnet with 5" clear bore and 9.25" height manufactured using NbTi wires generates an rms average field of 7.8\,T over the cavity volume at 81.0\,A.

The aperture and field distribution of the magnet determine the cavity dimension, 110\,mm in inner diameter and 215\,mm in inner height, which maximizes the scan rate.
The double-cell cavity comprised two identical copper pieces, each of which was fabricated as a single body with a 5mm-thick partition in the middle to form a two-cell structure when assembled.
A sapphire rod of 6.9\,mm diameter, capable of tuning the resonant frequency between 2.84 and 3.36\,GHz, was introduced in each cell.
The unloaded cavity $Q$-factors were measured using a network analyzer to be $\sim60,000$ over the tuning range.

\begin{figure}
    \centering
    \includegraphics[width=\linewidth]{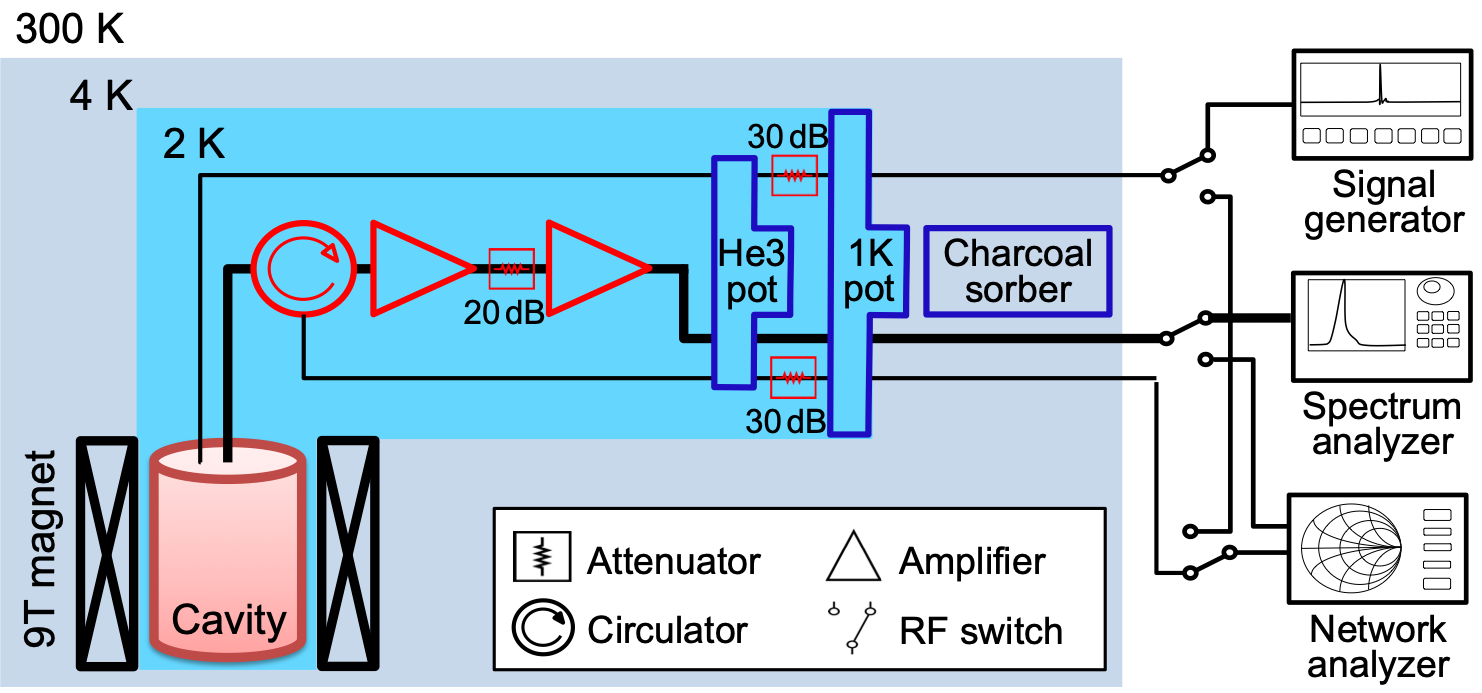}
    \caption{Schematic view of the experimental setup.}
    \label{fig:schematic}
\end{figure}

The major component of the receiver chain consists of two cryogenic high electron mobility transistors (HEMTs), thermally connected to the 1K pot, with a circulator placed in front and a 20dB attenuator in between.
The signal from the cavity is picked up by a strongly coupled antenna and amplified by the HEMT chain and further transferred through a series of RF cables to a spectrum analyzer (SA) at room temperature.
The total gain of the chain is $\sim55$\,dB, large enough to suppress the measured SA baseline noise of $4.8\times10^{-19}$\,W/Hz.
The noise temperature of the chain was measured to be $\sim1.5$\,K using the Y-factor method~\cite{note:Y-factor} by introducing a heat source at cryogenic temperature.
This value was confirmed by an independent measurement using a noise diode at room temperature, similarly to Ref.~\cite{paper:noise_meas}.
Along with the cavity temperature, this dominantly composed the system noise temperature of $\sim 3.8$\,K.
The experimental setup is schematically shown in Fig.~\ref{fig:schematic}.

The antenna coupling to the TM$_{010}$-like mode was set to $\beta\simeq2$ to maximize the search rate (see Eq.~\ref{eq:scan_rate}).
In addition to the resonant frequency and cavity $Q$-factor for this lowest mode, the antenna couplings to the higher TM$_{110}$-like mode was measured over the scanning process in order to estimate field localization.
The SA was configured to operate in a FFT-based mode with a 100\,Hz bandwidth small enough to resolve the axion signal ($\Delta\nu_a \approx 3$\,kHz) and  a 1\,MHz span large enough to cover the overall cavity response ($\Delta\nu_c \approx 150$\,kHz).
The sweep time was set to 10 seconds so the thermal noise obeys a Gaussian distribution.
Over a sweep time, a series of processes, including down-conversion, digitization, and Fourier transform, were performed by the SA internal functions to produce a single power spectrum.
The efficiency of the SA was determined to be $\sim60\%$ by comparing with the expected fluctuations in noise power acquired over the sweep time.
In order to identify the ambient RF interference, an additional SA with an antenna directly connected to the input port was brought near the experiment and ran simultaneously with the primary SA.
The tuning step, $\Delta\nu=50$\,kHz, was set to be smaller than the cavity bandwidth. 

We assume that axions solely make up our local galactic halo, i.e. $\rho_a=0.45$\,GeV/cm$^3$.
Targeting at the statistical significance of 5 standard deviations above the system noise for axion-photon couplings 10 times higher than the KSVZ model, it was estimated that a 19-day data acquisition (DAQ) would enable us to scan over a 220\,MHz frequency range between 3.14 and 3.36\,GHz (corresponding to an axion mass range between 13.0 and 13.9\,$\mu$eV).
To reach the target sensitivity, the required acquisition time is $\sim 5$ minutes and thus 30 power spectra were recorded at a fixed frequency.
This setup yielded $\delta P_{\rm sys}/\Delta\nu \approx 3.0\times10^{-25}$\,W/Hz using Eq.~\ref{eq:noise}.
The DAQ was interrupted every 6th record to measure the cavity properties, which resulted in a 4\% drop in DAQ efficiency.
Taking the SA efficiency into account, this yielded a scan rate of 4.4\,GHz/year, equivalent to 440\,kHz/year for a target sensitivity of the KSVZ coupling.

The acquired power spectra were processed following the basic analysis procedure described in Ref.~\cite{paper:HAYSTAC}.
For each raw spectrum, the power in the individual bins was divided by the baseline, obtained by fitting the spectrum using a five parameter function that reflects the physical parameters~\cite{paper:five_param_fit}, and subtracted by unity to produce power excess, denoted by $\delta$.
In the absence of signal or RF interference, the power excess density exhibited the standard normal distribution.
At every tuning step, 30 such processed spectra were combined to produce an average spectrum.
Spurious RF interference, typically characterized by a significant power excess within a single bin, was removed by considering the bin-by-bin correlation with the auxiliary data obtained by the secondary SA.
We imposed a criterion for the excess power, being $<5\sigma$ with an efficiency of 99.85\%.

For the convenience of further analysis, the power excess in each bin was re-scaled by multiplying by $(P_{\rm sig}/P_{\rm sys})^{-1}$, where $P_{\rm sig}$ is the signal power expected from KSVZ axions at the maximum peak of the signal distribution and weighted by the cavity response and $P_{\rm sys}$ is the system noise power obtained by $P_{\rm sys}=k_BT_{\rm sys}\Delta\nu$, so that all of the frequency bins had normalized power excess.
Over the course of tuning, a total of 4,336 re-scaled spectra with 50\,kHz interval were obtained.
These spectra were vertically combined bin by bin to construct a single broad spectrum spanning the entire frequency range.
The power excess in frequency bin $i$ of the combined spectrum was obtained from a weighted average over all contributions from individual 1MHz spectra~\cite{book:stat}.
Any single bins with a spurious peak exceeding 5$\sigma$ were additionally removed.
The weighted power excess of the combined spectrum, again in the absence of signal, followed the standard normal distribution.

Since the signal shape of the virialized axions has a finite bandwidth determined by its quality factor, $Q_a\approx10^6$~\cite{paper:line-shape}, combining neighboring bins belonging to the bandwidth statistically enhances the signal. 
At each frequency bin, the power of the neighboring bins was weighted by the expected signal power density distribution, given by the axion line-shape with the axion rest mass equal to the frequency, and summed up to be assigned to that bin.
The final spectrum, called grand spectrum, represents the maximum likelihood estimate that can be obtained from the axion model assuming the axion mass lies in each bin.
After the process, however, we observed that the power excess distribution became wider ($\sigma=1.31$).
We found a positive correlation between the nearest neighboring bins due to the so-called zero padded FFT, chosen as the SA analysis mode, which was verified by a Monte Carlo (MC) study.
The observed number was also confirmed by comparing baseline spectra from the SA with uncorrelated white noise spectra generated by MC.
This correlation induced a 14\% reduction in overall sensitivity.
Since the effect was understood, the width of the power distribution was renormalized to unity for convenience.

A null hypothesis on axions with ${\rm SNR}=5$ constructs a Gaussian signal power excess with $\mu=5$ and $\sigma=1$, and $3.718\sigma$ was set as the threshold for the 90\% confidence bound.
Among a total of 2,168,116 frequency bins, there were 36 candidates exceeding the threshold.
Additional data were acquired at the corresponding frequencies and added to the original data to be fully re-analyzed.
After three iterations of this re-scanning process, all the candidates remained below the threshold, indicating they were due to statistical fluctuations.
This number of candidates turned out to be statistically consistent with the result from a Gaussian white noise.

\begin{figure*}
    \centering
    \includegraphics[width=\textwidth]{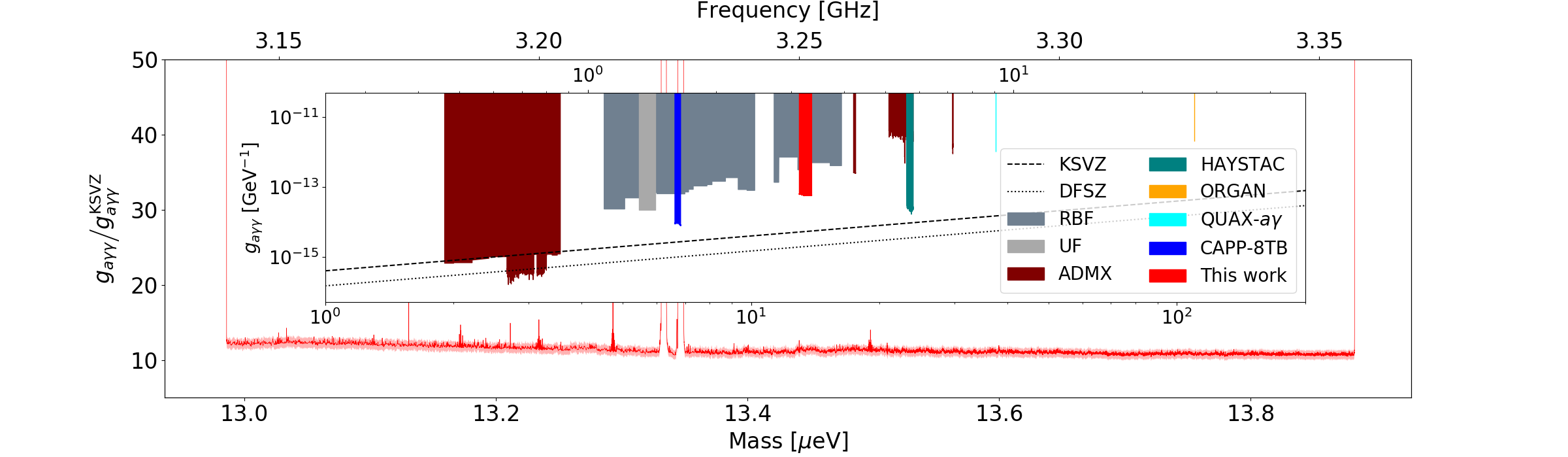}
    \caption{Exclusion limits on axion-photon coupling by CAPP-9T at 90\% CL.
    The large gaps at around 13.35\,$\mu$eV and small spikes correspond to the mode mixing regions and removed bins attributed to the spurious interference, respectively.
    The light red band represents the total uncertainty described in the text.
    %The inset compares our result with those from other cavity haloscope experiments~\cite{paper:RBF, paper:UF, paper:ADMX, paper:sidecar, paper:HAYSTAC, paper:QUAX-ag, paper:ORGAN, paper:CAPP-8TB} in a larger scale.}
    Results from other cavity experiments~\cite{paper:RBF, paper:UF, paper:ADMX, paper:sidecar, paper:HAYSTAC, paper:QUAX-ar, paper:ORGAN, paper:CAPP-8TB} are compared in the inset.}
    \label{fig:exclusion}
\end{figure*}

Since the power excess was normalized in the unit of the KSVZ axion signal during the re-scaling process, the final SNR becomes the reciprocal of the standard deviation in each bin of the grand spectrum.
Therefore, the axion-photon signal to be observed with SNR of 5 corresponds to the KSVZ coupling multiplied by $\sqrt{5\sigma}$, which determines the sensitivity to the KSVZ model.
From our data, we found $\sigma=24.8$ on average over the entire spectrum with no signal observed and hence we rejected the null hypothesis at 90\% confidence level (CL).
This excluded dark matter axions with coupling $g_{a\gamma\gamma} \gtrsim 11.1\times g_{a\gamma\gamma}^{\rm KSVZ}$ at the same CL in the mass range between 13.0 $\mu$eV and 13.9 $\mu$eV.
Figure~\ref{fig:exclusion} plots our exclusion limits, compared with other experimental results.

Several uncertainties were considered in the exclusion limits.
The noise figure of the receiver chain was measured independently using two noise sources -- heat source and noise diode.
The statistical uncertainty from the former, and the systematic uncertainty, quoted as the difference between the two, contributed to the uncertainty of noise measurement at the 5\% level.
The measured loaded $Q$-factors had typical fluctuations of $\sim400$, yielding a relative uncertainty of $\sim2\%$.
The form factors were evaluated to be $\sim0.55$ based on simulation and corrected by the degradation estimated from Fig.~\ref{fig:gap_opt} and Fig.~\ref{fig:misalignment}.
The former yielded the relative value of $0.97^{+0.01}_{-0.02}$ at the optimal center gap, while the latter returned the correction factor of $0.995\pm0.001$ from the measured reflection coefficients of $0.03\pm0.01$\,dB.
%The combination of the two was reflected in the form factor estimation.
The total uncertainty was obtained by a quadratic sum of these individual contributions.

In conclusion, we addressed the unique characteristics of the multiple-cell cavity design and its high tolerance against construction inaccuracy and misalignment, which proved it is effectively applicable for high-mass axion signal searches.
The first experiment to demonstrate this cavity concept was performed by employing a double-cell cavity, looking for dark matter axions with mass $>10\,\mu$eV and statistical significance of $5\sigma$.
Compared to a conventional single cavity, the higher performance with high reliability, as evident by Figs.~\ref{fig:gap_opt} and~\ref{fig:misalignment}, enabled a fast scan over $>200$\,MHz above 3\,GHz.
This result exceeds previous ones within the mass range $13.0-13.9\,\mu$eV and sets new exclusion limits for axion coupling to photon of $g_{a\gamma\gamma} \gtrsim 11.1\times g_{a\gamma\gamma}^{\rm KSVZ}$ at 90\% CL.
This confirms that this novel cavity design can contribute considerably to the next-generation invisible dark matter axion searches.

\begin{acknowledgments}
This work was supported by the Institute for Basic Science (IBS-R017-D1-2020-a00/ IBS-R017-Y1-2020-a00).
J. E Kim is also supported in part by the National Research Foundation grants NRF-2018R1A2A3074631.
\end{acknowledgments}

\end{document}